\documentclass[final]{article}

\usepackage[english]{babel}
\usepackage{url}
\usepackage{amsfonts,amsmath,amssymb,mathrsfs,amsthm}
\usepackage{dsfont}
\usepackage{graphicx}
\usepackage{color}
\usepackage{setspace}

\newtheorem{theorem}{Theorem}[section]
\newtheorem{lemma}[theorem]{Lemma}
\newtheorem{definition}[theorem]{Lemma}
\newtheorem{remark}[theorem]{Remark}

\usepackage{authblk}

\begin{document}

\title{THE QUANTUM MARGINAL PROBLEM}

\author{Christian Schilling\footnote{comments to \textbf{christian.schilling@physics.ox.ac.uk}}}
\affil{\small Clarendon Laboratory, University of Oxford, Oxford OX1 3PU, United Kingdom}

\maketitle

\begin{abstract}
The question of whether given density operators for subsystems
of a multipartite quantum system are compatible to one common total density operator is known as
the quantum marginal problem. We briefly review the solution of a subclass of such
problems found just recently. In particular, this provides the solution of the
$1$-body $N$-representability problem. Its solution, the so-called generalized
Pauli constraints, restrict the set of mathematically possible fermionic occupation
numbers significantly, and strengthens Pauli's exclusion principle. Moreover, we review
the study of a concrete physical model of interacting fermions confined to a
harmonic trap. There, we found occupation numbers close, but not exactly on the
boundary of the allowed region. This new effect of quasipinning is physically relevant
since it corresponds to a simplified structure of the corresponding $N$-fermion quantum state.
\end{abstract}


\section{Introduction}\label{sec:intro}
Since most quantum effects emerge from the interaction between two or more quantum systems
as e.g.~the interaction of a system with an environment or the interaction of macroscopically
many electrons the concept of a multipartite quantum system is fundamental. We describe the
quantum state of such a system $\mathcal{J}$ built up from subsystems $A, B, C, \ldots$ by a
density operator $\rho_{\mathcal{J}}$ acting on the Hilbert space $\mathcal{H}_{\mathcal{J}} =
\mathcal{H}_{\mathcal{A}}\otimes \mathcal{H}_{\mathcal{B}}\otimes \ldots$, the tensor product of
the separable Hilbert spaces of the subsystems $A,B,\ldots$. To describe properties of
some subsystem $\mathcal{I}$ of $\mathcal{J}$, as e.g.~$B$ or $AC$ (containing system $A$ and $C$),
it suffices to deal with the corresponding reduced density operator (marginal) of system $\mathcal{I}$
\begin{equation}\label{partialtrace}
\rho_\mathcal{I} = \mbox{Tr}_{\mathcal{J}\setminus \mathcal{I}}[\rho_{\mathcal{J}}],
\end{equation}
obtained by tracing out the complementary system $\mathcal{J}\setminus \mathcal{I}$ of $\mathcal{I}$.
According to Eq.~(\ref{partialtrace})
it is clear that marginals arising from the same total quantum state $\rho_{\mathcal{J}}$ need to
fulfil certain compatibility conditions. This gives rise to the quantum marginal problem (QMP)
\begin{definition}[quantum marginal problem]\label{def:qmp}
For a given family  $\mathcal{K}$ of subsystems $\mathcal{I}$ of $\mathcal{J}$ the
quantum marginal problem $\mathcal{M}_{\mathcal{K}}$ is the problem of determining and describing
the set $ \Sigma_{\mathcal{K}}$ of tuples $(\rho_{\mathcal{I}})_{\mathcal{I} \in \mathcal{K}}$ of
compatible marginals. Compatible here means that there exists a density operator $\rho_{\mathcal{J}}$
for the total system such that $\forall \mathcal{I} \in \mathcal{K}$
\begin{equation}
\rho_{\mathcal{I}}= \mbox{Tr}_{\mathcal{J} \setminus \mathcal{I}}[\rho_{\mathcal{J}}]\,.
\end{equation}
\end{definition}
In this work, due to mathematical reasons, we assume $\mathcal{H}_{\mathcal{J}}$ to be finite dimensional.
Moreover, we may think of $\mathcal{I} \subset  \mathcal{J}$ as one particle, a few particles, a system of macroscopically
many particles or as just the spin degree of freedom of a single electron.
\begin{figure}[h]
\centering
\includegraphics[scale=0.50]{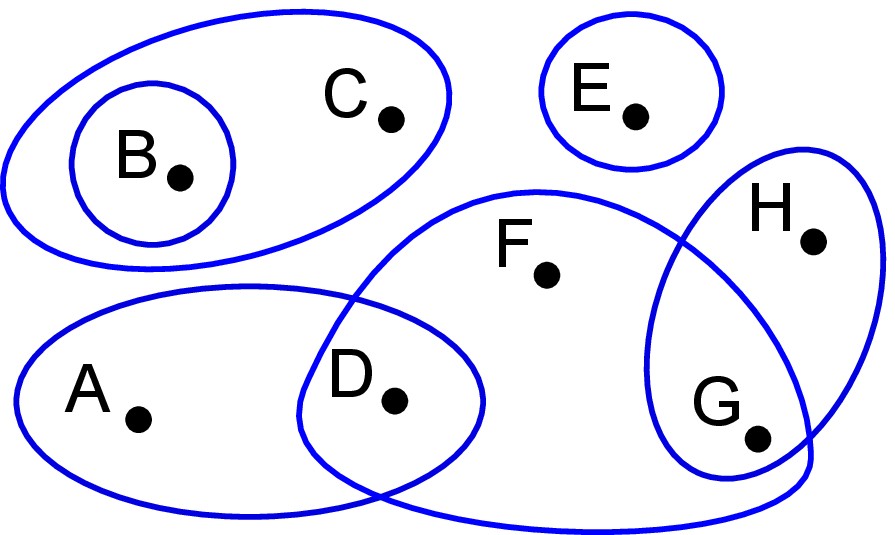}\hspace{1.2cm}
\includegraphics[scale=0.50]{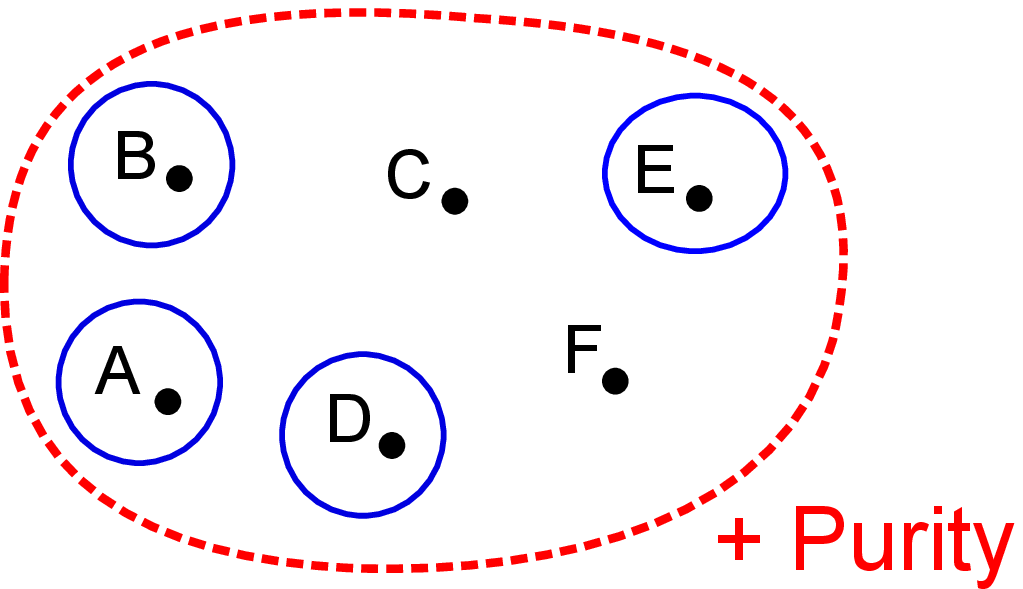}
\caption{Illustration of the quantum marginal problem in its general form (left) and its pure univariant form (right). See also text.}
\label{fig:qmp}
\end{figure}
The QMP is also illustrated in Fig.~\ref{fig:qmp}. There, every physical system $I = A,B,C,\ldots$ is symbolically described by a black dot.
The family $\mathcal{K}$ of systems of interest is illustrated by `blue islands', where every island describes one subset $\mathcal{I} \in \mathcal{K}$. In addition, one may modify the QMP by exposing restrictions on the total state as e.g.~purity or if all the subsystems are identical a fermionic or bosonic exchange symmetry.

One of the most important QMP is given by (see \cite{Col})
\begin{definition}[$r$-body $N$-representability problem]\label{def:nrep}
Given a system of $N$ identical fermions, with corresponding Hilbert space $\mathcal{H}_N^{(f)} = \wedge^N[\mathcal{H}_1^{(d)}]$ of antisymmetric states, where the dimension $d$ of the $1$-particle Hilbert $\mathcal{H}_1^{(d)}$ may be infinite. For fixed $r \in \{1,2,\ldots,N-1\}$ the problem of determining the set $D_r^{(p/e)}$ of possible $r$-reduced density operators ($r$-RDO) $\rho_r$ arising via partial trace from a corresponding pure/ensemble $N$-fermion density operator $\rho_N$ on $\mathcal{H}_N^{(f)}$ is the \emph{$r$-body pure/ensemble-$N$-representability problem}.
\end{definition}

The paper is arranged as follows. In the next section we present the solution of the subclass
of pure univariant QMP and give the reader an idea how to derive it.
In Sec.~\ref{sec:gpc} we focus on the $1$-body pure $N$-representability problem and explain
that the antisymmetry of the $N$-fermion state implies so-called generalized Pauli constraints,
restrictions of fermionic occupation numbers stronger than Pauli's exclusion principle. Their
role for ground states is studied in Sec.~\ref{sec:model} by solving a model of few
fermions confined by a harmonic trap. We find the effect of quasipinning.
In the last section, Sec.~\ref{sec:physrev}, we briefly discuss the physical relevance of pinning
and explain why this may lead to a generalized Hartree-Fock method.

\section{Pure Univariant Quantum Marginal Problem and Solution}\label{sec:qmp}
Solving the QMP in its general form is practically impossible. This was shown in
\cite{Liu} for the $2$-body $N$-representability problem, by proving that it
belongs to the quantum Mermin-Arthur complexity class (a generalization of the NP class).
As a consequence one was tempted to focus on a subclass of the QMP, the so-called pure
univariant QMP. There, the family $\mathcal{K}$ (recall Definition \ref{def:qmp}) is given
by disjoint subsystems $\mathcal{I}\subset\mathcal{J}$ and the total state is
required to be pure. This is illustrated on the right side of Fig.~\ref{fig:qmp}.
There, all blue islands are non-overlapping (univariant) and for each island we collect all
its black dots by one. This restriction of non-overlapping `blue islands' leads to a significant
simplification of the corresponding QMP:
\begin{remark}\label{rem:uniteq}
Due to the unitary equivalence
\begin{equation}
(\rho_{\mathcal{I}})_{\mathcal{I} \in \mathcal{K}} \,\mbox{compatible} \Rightarrow (U_{\mathcal{I}}
\rho_{\mathcal{I}}U_{\mathcal{I}}^\dagger)_{\mathcal{I} \in \mathcal{K}} \,\mbox{compatible}\,,\,\,\,\forall
\,\mbox{unitaries}\,U_{\mathcal{I}}\, \mbox{on}\,\mathcal{H}_{\mathcal{I}}
\end{equation}
the set $\Sigma_{\mathcal{K}}$ of compatible marginals for univariant QMP $\mathcal{M}_{\mathcal{K}}$
is described by conditions on their spectra $\vec{\lambda}_{\mathcal{I}}$, only.
\end{remark}

In 2004, Klyachko \cite{Kly4} has solved the pure univariant QMP. Further impotant contributions came from
Daftuar and Hayden \cite{Daft} and Christandl and Mitchison \cite{MC}. We present the abstract solution and review briefly the main
idea for deriving it.

\begin{theorem}
Let $\mathcal{K}$ define a pure univariant QMP $\mathcal{M}_{\mathcal{K}}$ (recall Definition \ref{def:qmp}).
According to Remark \ref{rem:uniteq} the set $\Sigma_{\mathcal{K}}$ of compatible marginals is described by their
decreasingly-ordered spectra $\vec{\lambda}_{\mathcal{I}}$. The set of possible spectra
$\vec{\lambda} \equiv (\vec{\lambda}_{\mathcal{I}})_{\mathcal{I}\in \mathcal{K}}$ forms a polytope
$\mathcal{P}\subset \mathbb{R}^d$. It highly depends on the dimensions of all local Hilbert spaces
$\mathcal{H}_A$, $\mathcal{H}_B$,\ldots and $d$ is the corresponding dimension of those $\vec{\lambda}$-vectors.
\end{theorem}
Since a polytope is nothing else but an intersection of finitely many Euclidean half spaces the set of possible
vectors $\vec{\lambda}$ is described by a finite family of so-called marginal constraints, linear conditions
on the eigenvalues $ \vec{\lambda}$. Klyachko \cite{Kly4} provides an
(quite abstract) algorithm for calculation those constraints.

\subsection{Derivation of marginal constraints}
In this section we give the reader an idea how to derive marginal constraints and follow
quite closely \cite{Daft}.
For this it is instructive to study an elementary prototype of a QMP, the spectral version of
$\mathcal{M}_{A,AB}$. It asks when spectra $\vec{\lambda}_A \in \mathbb{R}^{d_A}$, $\vec{\lambda}_{AB}\in \mathbb{R}^{d_A d_B}$ are compatible
in the sense that there exists a total state $\rho_{AB}$ with spectrum $\vec{\lambda}_{AB}$ such that
its marginal $\rho_A$ has spectrum $\vec{\lambda}_A$.

First, we need to introduce the concept of a flag induced by a hermitian operator, the complex
Grassmanian, its Schubert cells and a variational principle due to Hersch and Zwahlen.
\begin{definition}\label{defflag}
Let $\mathcal{H}$ be a $d$-dimensional complex Hilbert space. A \emph{complete flag} $F_{\bullet}$ is a maximal
sequence of nested linear subspaces, i.e.
\begin{equation}
F_{\bullet}:=[{0}= F_0 \lneq F_1 \lneq \ldots \lneq F_{d-1}\lneq F_d = \mathcal{H}]
\end{equation}
\end{definition}
In particular, complete flags can be induced by non-degenerate hermitian operators according
\begin{definition}\label{flagoperator}
Given a hermitian operator $A$ on a $d$-dimensional Hilbert space with non-degenerate spectrum
$a=(a_1,\ldots,a_d)$ arranged in decreasing order. $A$ then induces a complete flag $F_{\bullet}(A)$
according
\begin{equation}
F_{i}(A) = \langle v_1,\ldots, v_i\rangle\qquad,\, \forall i=0,1,\ldots,d\,,
\end{equation}
where $v_j$ is the eigenvector corresponding to the eigenvalue $a_j$ and $\langle \cdot\rangle$
denotes the span of vectors.
\end{definition}
\begin{definition}\label{SchubertcellGr}
Let $\mathcal{H}$ be a $d$-dimensional Hilbert space and $F_{\bullet}$ a complete flag. Then for
every binary sequence $\pi \in \{0,1\}^d$ we define the Grassmannian Schubert cell
$S_{\pi}^{\circ}(F_{\bullet})$ by
\begin{equation}
S_{\pi}^{\circ}(F_{\bullet}):= \{ V\leq\mathcal{H}\,|\, \forall i=1,\ldots, d: \mbox{dim}((V\cap F_i)
/(V \cap F_{i-1}))= \pi_i \}
\end{equation}
These cells are subsets of the Grassmannian $\mbox{Gr}_{\|\pi\|_1,d}$, which are defined as
\begin{eqnarray}\label{GrFlagPredefinition}
\mbox{Gr}_{\|\pi\|_1,d} &:=& \{V\leq \mathcal{H} \,|\, \mbox{dim}(V)= \|\pi\|_1\}
\end{eqnarray}
and $\|\pi\|_1 \equiv \sum_{k=1}^d \pi_k$\,.
\end{definition}
\begin{remark}
The binary sequence $\pi$ defines the indices at which the components (vector spaces) of the sequence
$V\cap F_0 \leq V\cap F_1 \leq \ldots \leq V\cap F_d$ increase their dimension. The label
$^{\circ}$ indicates that the Schubert cells are open w.r.t.~the natural topology. The closures of
these Schubert cells are called Schubert varieties.
Moreover, it is well known (see e.g.~\cite{CSthesis}) that $\mbox{Gr}_{k,d}$ is
a projective algebraic variety and the Schubert varieties form subvarieties.
\end{remark}

Now, we can express sums of \emph{arbitrary} eigenvalues of a hermitian operator by a variational
principle:
\begin{lemma}[Hersch-Zwahlen]\label{lem:HerschZwahlen} Let $\rho$ be a hermitian operator with non-
degenerate spectrum $\vec{\lambda}$ arranged in decreasing order, $\pi \in \{0,1\}^d$ a binary sequence
of length $d$. Then
\begin{equation}
\sum_{j=1}^d \pi_j \lambda_j = \min \limits_{V \in S_{\pi}^{\circ}(\rho)}(\mbox{Tr}[P_V \rho])\,,
\end{equation}
where $P_V$ is the orthogonal projection operator onto the subspace $V$.
\end{lemma}
The proof is elementary and can e.g.~be found in \cite{Daft,CSthesis}.
Lemma \ref{lem:HerschZwahlen} can be used to derive necessary conditions on compatible spectra
$\vec{\lambda}_A,\vec{\lambda}_{AB}$. For this, we denote their total state by $\rho_{AB}$ with
$\rho_A = \mbox{Tr}_B[\rho_{AB}]$ and choose arbitrary binary sequences $\pi\in \{0,1\}^{d_A}$
and $\sigma\in \{0,1\}^{d_A d_B}$. Then, whenever the intersection property
(the \emph{dual binary sequence} $\hat{\sigma}$ of a sequence $\sigma$ of length $d$ is defined by $\hat{\sigma}_k := \sigma_{d-k+1}$)
\begin{equation}\label{intersectionprop1}
\left(S_{\pi}^{\circ}(\rho_A) \otimes \mathcal{H}^{(B)} \right) \cap S_{\hat \sigma}^{\circ}(\rho_{AB}) \neq \emptyset
\end{equation}
holds, we obtain
\begin{eqnarray}\label{deviationspecineq1}
\lefteqn{\sum_{j=1}^{d_A}\pi_j \lambda_j^{(A)} - \sum_{i=1}^{d_A d_B} \sigma_i \lambda_i^{(AB)}}&&\nonumber \\
&=& \sum_{j=1}^{d_A} \pi_j \lambda_j^{(A)} + \sum_{i=1}^{d_A d_B} \sigma_i (-\lambda_i^{(AB)}) \nonumber \\
&=& \min \limits_{V \in S_{\pi}^{\circ}(\rho_A)}(\mbox{Tr}_A[P_V \rho_A]) + \min \limits_{W \in S_{\hat \sigma}^{\circ}(-\rho_{AB})}(\mbox{Tr}_{AB}[P_W (-\rho_{AB})]) \nonumber \\
&=& \min \limits_{V\otimes \mathcal{H}^{(B)} \in S_{\pi}^{\circ}(\rho_A)\otimes \mathcal{H}^{(B)}}(\mbox{Tr}_{AB}[P_{V\otimes \mathcal{H}^{(B)}} \rho_{AB}]) + \min \limits_{W \in S_{\hat \sigma}^{\circ}(-\rho_{AB})}(\mbox{Tr}_{AB}[P_W (-\rho_{AB})]) \nonumber \\
&\leq& \mbox{Tr}_{AB}[P_{W_0} \rho_{AB}] + \mbox{Tr}_{AB}[P_{W_0} (-\rho_{AB})] \nonumber \\
&=&0 \,,
\end{eqnarray}
where we applied $S_{\sigma}^{\circ}(-\rho) = S_{\hat{\sigma}}^{\circ}(\rho) $ in the third line and
(\ref{intersectionprop1}) was used in the second last line, with an element $W_0 \in  \left(S_{\pi}^{\circ}(\rho_A) \otimes \mathcal{H}^{(B)}\right) \cap S_{\hat \sigma}^{\circ}(\rho_{AB})$.
Hence, we obtain a spectral inequality
\begin{equation}\label{spectralineq1}
\boxed{
\sum_{j=1}^{d_A}\pi_j \lambda_j^{(A)} \leq \sum_{i=1}^{d_A d_B} \sigma_i \lambda_i^{(AB)}} \,.
\end{equation}

Most of the work by Klyachko, Daftuar and Hayden concerns the
intersection property Eq.~(\ref{intersectionprop1}). Using cohomology theory allows to map
it to an algebraic level and study it
there systematically. The structure of Schubert varieties as projective algebraic
varieties is essential for that. It gives them an algebraic meaning, since they turn out
to stand in a one-to-one correspondence with the generators of the cohomology ring of the
complex Grassmannian (see e.g.~\cite{Daft}).

In addition, Klyachko did not only provide an algorithm for calculating these necessary marginal constraints, but
also proved that they are sufficient \cite{Kly4} for the compatibility of spectra/marginals.

\section{Generalized Pauli Constraints}\label{sec:gpc}
In this section we emphasize that the solution of the pure univariant QMP implies
a generalized (and stronger) Pauli exclusion principle.
\begin{figure}[h]
\centering
\includegraphics[scale=0.7]{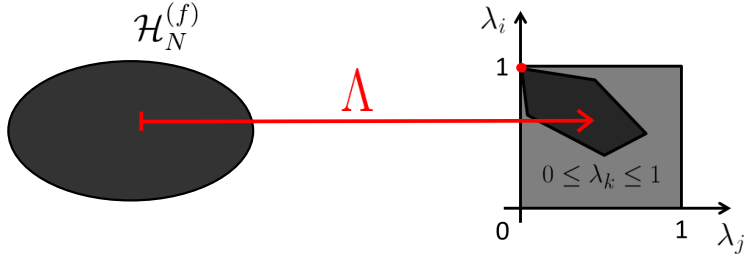}
\caption{The family of antisymmetric $N-$particle states maps to the family of possible natural occupation numbers $\vec{\lambda}$, which turns out to be a proper subset of the Pauli hyper cube. The Hartree-Fock point is shown as red dot.}
\label{fig:fqmp}
\end{figure}
For this we introduce the map
\begin{eqnarray}
\Lambda:&\wedge^N[\mathcal{H}_1^{(d)}]& \qquad \rightarrow \qquad \mathbb{R}^d \nonumber \\
& |\Psi_N\rangle & \qquad \mapsto \qquad  \vec{\lambda} \equiv \mbox{spec}\left(N \mbox{Tr}_{N-1}[|\Psi_N\rangle\langle\Psi_N|]\right)\,.
\end{eqnarray}
$\Lambda$ maps pure antisymmetric quantum states via their $1$-particle density operator $\rho_1\equiv N
\mbox{Tr}_{N-1}[|\Psi\rangle \langle\Psi_N|]$ (trace-normalized to the particle number $N$) to its decreasingly-ordered
\emph{natural occupation numbers} (NON) $\vec{\lambda} \equiv (\lambda_1,\ldots,\lambda_d)$. Pauli's exclusion principle can then be formulated as
\begin{equation}
0\leq \lambda_i \leq 1\,,
\end{equation}
which is a strong restriction of the range of $\Lambda$. This well-known implication of
Pauli's exclusion principle from the antisymmetry is also illustrated in Fig.~\ref{fig:fqmp}.
A natural question arises: Are there further restrictions on fermionic NON $\vec{\lambda}$? This question is equivalent to the $1$-body pure $N$-representability
problem (see Definition \ref{def:nrep} and Remark \ref{rem:uniteq}). Moreover, since this is a pure
univariant QMP the answer to that question is `yes'. Klyachko \cite{Kly2,Kly3} provides an algorithm
that allows to calculated these so-called generalized Pauli constraints for each fixed $N$ and $d$.
They all take the form of linear inequalities,
\begin{equation}\label{gPC}
D_i^{(N,d)}(\vec{\lambda})=\kappa_i^{(0)}  + \kappa_i^{(1)} \lambda_1+\ldots+\kappa_i^{(d)} \lambda_{d} \geq 0\,,
\end{equation}
with affine coefficients $\kappa_i^{(j)} \in \mathbb{Z}$, $j=0,1,\ldots,d$ and $i=1,2,\ldots,r^{(N,d)}$. The
number $r^{(N,d)}$ of such constraints (\ref{gPC}) increases drastically with $d$.
To give the reader an idea how non-trivial they are we consider the setting $\wedge^3[\mathcal{H}_1^{(6)}]$.
Already in 1972, Borland and Dennis found the necessary and sufficient conditions on $\vec{\lambda}$.
They are given by \cite{Borl1972}
\begin{eqnarray}
&&\lambda_1+\lambda_6 = \lambda_2+\lambda_5 = \lambda_3+\lambda_4 = 1\,, \label{d=6a} \\
&&D^{(3,6)}(\vec{\lambda}) \equiv 2-(\lambda_1 +\lambda_2+\lambda_4) \geq 0 \label{d=6b} \,,
\end{eqnarray}
where the NON are always ordered decreasingly, $\lambda_1\geq\lambda_2\geq\ldots \geq\lambda_6\geq 0$.
Notice that the inequality $D^{(3,6)}(\vec{\lambda}) \geq 0$ is manifestly stronger than Pauli's exclusion principle, which just states
that $2-(\lambda_1 +\lambda_2)\geq 0$. That some constraints take the form of equalities (instead of inequalities) is specific
and happens only for this small setting of three fermions and a $6$-dimensional $1$-particle Hilbert space.

It is important to notice that the existence of that polytope $\mathcal{P}_{N,d}$ and the
corresponding restriction of fermionic occupation numbers is purely kinematic and not
related to any Hamiltonian. To use this beautiful new mathematical structure revealed
by Klyachko for physics the first task is to understand where the occupation numbers
of relevant fermionic quantum states do lie. If we consider e.g.~ground states of non-interacting fermions confined by some
external potential the position of the corresponding $\vec{\lambda}$-vector is obvious. The ground state
$|\Psi_0\rangle$ is given by a single Slater determinant $|1,2,\ldots,N\rangle$, the antisymmetrized tensor product of
the $1$-particle states $|i\rangle$ corresponding to the lowest $N$ $1$-particle energy levels of the external trap.
This ground state yields the NON $\vec{\lambda}=(1,\ldots,1,0,\ldots)$ (red dot in Fig.~\ref{fig:fqmp}). If we turn on some interaction with coupling strength
$\kappa$ the NON $\vec{\lambda}(\kappa)$ will move away from the so-called Hartree-Fock point $(1,\ldots,1,0,\ldots)$.
The central question is then whether it moves towards the middle of the polytope or whether it still lies on the boundary
of the polytope. In the latter case we say that the NON are \emph{pinned to the boundary} of the polytope. Before investigating
this question we explain why possible pinning may be interesting.
\begin{figure}[h!]
\centering
\includegraphics[width=2.3cm]{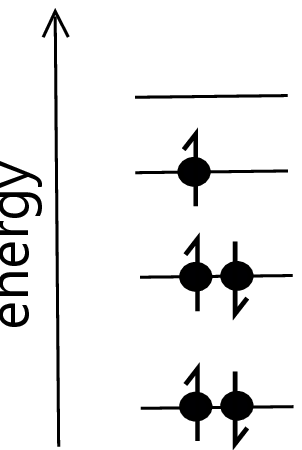}\hspace{2.0cm}
\includegraphics[width=3.5cm]{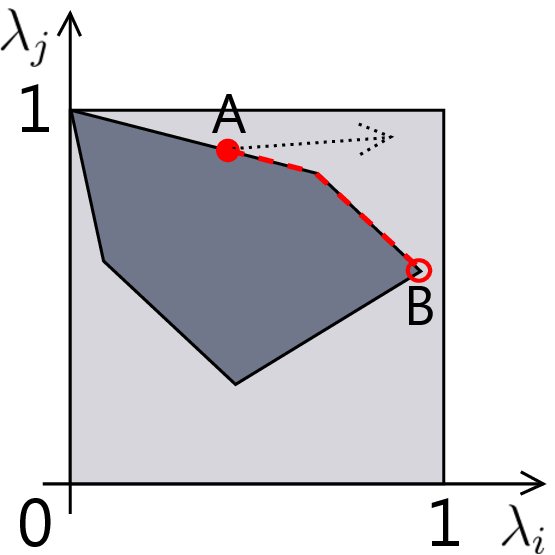}
\centering
\caption{Right: Initial NONs on the boundary (A) of the polytope. Any time evolution that would like to drive them out of the polytope (dashed arrow) is dominated by the geometry of the polytope rather than by the Hamiltonian. This kinematical influence by the generalized Pauli constraints generalizes that by the Pauli exclusion principle (on the left). There e.g.~the electron in the highest shell cannot decay to a lower one.}
\label{fig:PCpinned}
\end{figure}
The first point was suggested by Klyachko in \cite{Kly1}, where he also introduced the effect of pinning.
\begin{enumerate}
\item Given an $N$-fermion Hamiltonian $H_N$. Its ground state $|\Psi_N\rangle$ can be obtained via a minimization of the energy expectation value
$E[\Psi_N] \equiv \langle\Psi_N|H_N|\Psi_N\rangle$. If the ground state turns out to be pinned to the boundary of the polytope a generalized Pauli constraint is active for the minimization in the sense that any further minimization of the energy would violate it. In that case the corresponding constraint would  have a strong influence on the ground state and its energy.
\item Consider an $N$-fermion system initially prepared in a quantum state $|\Psi_N\rangle$ with NON $\vec{\lambda}$ pinned to the boundary $\partial\mathcal{P}_{N,d}$ of the polytope. By switching on a unitary time evolution for that system the NON $\vec{\lambda}(t)$ will begin to move. For some very specific time evolutions $\vec{\lambda}(t)$ may like to leave the polytope. Such a scenario is illustrated on the right side of Fig.~\ref{fig:PCpinned}. Since $\vec{\lambda}(t)$ cannot leave the polytope, it will move along the boundary from the initial point $A$ to the final point $B$. In that case the time evolution is dominated by the geometry of the polytope rather than by the Hamiltonian. This kinematical effect on time evolutions is a generalization of a similar more elementary effect shown on the left side of Fig.~\ref{fig:PCpinned}. There, we can see some (non-interacting) electrons occupying low-lying energy shells. Coupling this systems to photons may in principle lead to a decaying of the electrons in the higher energy shells to the lowest one. However, this is impossible due to Pauli's exclusion principle. In the same way the physics of solid bodies at low temperatures is dominated by the electrons close to the Fermi level.
\end{enumerate}

\section{Quasipinning for $N$-Harmonium}\label{sec:model}
To understand how ground states of interacting fermions look like from the new viewpoint of generalized
Pauli constraints we studied a model of few harmonically coupled spinless fermions in one dimension confined
by a harmonic trap. For details of this study we refer to \cite{CS2013,CS2013NO,CSthesis}. The Hamiltonian reads
\begin{equation}\label{Hamiltonian}
H = \sum_{i=1}^N\,\left(\,\frac{p_i^2}{2m}+\frac{1}{2}m\omega^2 x_i^2\,\right) + \frac{1}{2} K\,\sum_{i,j=1}^N \, (x_i - x_j)^2
\end{equation}
and acts on the fermionic Hilbert space $\mathcal{H}_N^{(f)} \equiv \wedge^N[\mathcal{H}_1]$, where the $1$-particle Hilbert space
is given by $\mathcal{H}_1 = L^2(\mathbb{R})$.
Its ground state can easily be found (see e.g.\ \cite{harmOsc2012}),
\begin{equation}\label{groundstate}
\Psi_N(\vec{x}) = c_0\,\times\, \prod_{1\leq i<j\leq N}(x_i-x_j) \times \exp\left[- c_1 (x_1+\ldots+x_N)^2 - c_2 \vec{x}^2\right]
\end{equation}
and has some similarity to the famous Laughlin wave functions \cite{Laughlin1983}. We determined analytically the corresponding $1-$RDO
depending on the relative interaction strength $\kappa \equiv \frac{N K}{m\omega^2}$ for arbitrary $N$. For $N =3$ we calculated
analytically with high effort by applying degenerate Rayleigh-Schr\"odinger perturbation theory its eigenvalues $\lambda_i(\kappa)$ for the
regime of not too strong interaction $\kappa$. For the ground state we found for the smallest distance $D(\kappa)$ of the spectrum
$\vec{\lambda}(\kappa)$ to the polytope boundary \footnote{The analysis of the distance to the boundary is quite subtle. The underlying
$1$-particle Hilbert space is infinite-dimensional, but the polytopes are known only for dimensions $d\leq 10$. However, it turns out
that all except the largest seven NON are very small and a mathematical result presented in \cite{SuplMat} allows to neglect them and truncate
the spectrum to the first few NON.}
\begin{equation}\label{DHarmonium}
D(\kappa) \sim const\times \kappa^8 \,.
\end{equation}
We called this surprising behavior, to be not on the boundary but very close to it, \emph{quasipinning}. This effect is non-trivial
in the sense that the distance (\ref{DHarmonium}) to the polytope boundary $\partial P$ is by four orders in $\kappa$ smaller
than the distance to the Hartree-Fock point (lies on the boundary, see Figure \ref{fig:fqmp}), which behaves as $\kappa^4$. Moreover, quasipinning is not only present in the regime of week interaction ($|\kappa|$ small), but also for medium interaction strengths. E.g.\ for $\kappa \equiv \frac{3K}{m \omega^2} = \frac{1}{3}$ we found $D = 5.8\cdot10^{-8}$.

It is one of the open problems to explore whether few-fermion ground states exhibit generic quasipinning for not too strong
interactions in the sense that it is independent of the concrete interaction form. Moreover, the mechanism behind it is not clear
yet. Since quasipinning is weaker for the first few excited $N$-fermion states \cite{CSthesis} and vanishes for higher excitations
a promising candidate for this mechanism would be the conflict of the antisymmetry (implying the generalized Pauli constraints)
and the energy minimization:
If one skipped the antisymmetry for the $N$-particle quantum state one would find much lower ground state energies.
\section{Physical Relevance of Pinning}\label{sec:physrev}
In this section we explain that pinning as an effect in the $1$-particle picture allows to reconstruct the structure
of the corresponding $N$-fermion quantum state.

First, we introduce some notation. Given an $N$-fermion pure state $|\Psi_N\rangle \in \wedge^N[\mathcal{H}_1^{(d)}]$.
Its $1$-RDO $\rho_1 = N \mbox{Tr}_{N-1}[|\Psi_N\rangle\langle\Psi_N|]$ can be diagonalized according to
\begin{equation}
\rho_1 \equiv \sum_{k=1}^d\,\lambda_k\,|k\rangle\langle k|
\end{equation}
with decreasingly-ordered $\lambda_k$, the occupation numbers w.r.t.~the \emph{natural orbitals} (NO)
$|k\rangle$. The NO define an orthonormal basis $\mathcal{B}_1$ for $\mathcal{H}_1^{(d)}$, which induces an orthonormal basis for
$\mathcal{H}_N^{(f)}$, the Slater determinants $|\textbf{k}\rangle$, $\textbf{k}\equiv (k_1,\ldots,k_N)$ with $1\leq k_1 <\ldots
<k_N\leq d$. Moreover, we denote the fermionic creation and annihilation operators w.r.t.~$\mathcal{B}_1$ by $a_k^\dagger$ and $a_k$,
respectively.
%
%

Now, we consider a generalized Pauli constraint (\ref{gPC}) and define (for a fixed $|\Psi_N\rangle$ with NON $\vec{\lambda}$)
\begin{equation}\label{Dop}
\hat{D}:= \kappa^{(0)} \textbf{1} + \kappa^{(1)} a_1^{\dagger} a_1+\ldots+\kappa^{(d)} a_{d}^{\dagger}a_{d}\,.
\end{equation}
Since $\kappa_i \in \mathbb{Z}$ we have $\mbox{spec}(\hat{D}) \subset \mathbb{Z}$.
One can prove (see e.g.~\cite{Alexthesis}) the important result
\begin{lemma}\label{lem:Deigen}
For given generalized Pauli constraint $D(\cdot)\geq 0$ (see also Eq.~(\ref{gPC})) and fixed state $|\Psi_N\rangle$ with NON $\vec{\lambda}$ define $\hat{D}$ according to Eq.~(\ref{Dop}). Then, when $\vec{\lambda}$ is pinned by $D$, $D(\vec{\lambda})=0$, it follows that
\begin{equation}
\hat{D}|\Psi_N\rangle = 0\,.
\end{equation}
\end{lemma}
From Lemma \ref{lem:Deigen} we can immediately conclude that whenever NON are pinned to some facet of the polytope the corresponding $|\Psi_N\rangle$ has weight only in the $0$-eigenspace of the corresponding $\hat{D}$-operator (\ref{Dop}). By expanding $|\Psi_N\rangle$ w.r.t.~$\mathcal{B}_N$,
\begin{equation}
|\Psi_N\rangle = \sum_{\textbf{i}}\,c_{\textbf{i}}\,|\textbf{i}\rangle,
\end{equation}
we find a selection rule (recall (\ref{Dop})) due to Klyachko \cite{Kly1},
\begin{equation}\label{selec}
\hat{D} |\textbf{i}\rangle \neq 0 \qquad \Rightarrow \qquad c_{\textbf{i}} = 0\,.
\end{equation}
To emphasize the importance of (\ref{selec}) we apply the selection rule to an example.
Consider a state $|\Psi_3\rangle \in \wedge^3[\mathcal{H}_1^{(6)}]$ with NON $\vec{\lambda}$. The generalized Pauli constraints are given by (\ref{d=6a}) and (\ref{d=6b}). The first three constraints take independent of $\vec{\lambda}$ the form of equalities. According to Eq.~(\ref{selec}) this leads to universal structural implications for any arbitrary $|\Psi_3\rangle \in \wedge^3[\mathcal{H}_1^{(6)}]$.
In the expansion
\begin{equation}
|\Psi_3\rangle = \sum_{1\leq i_1 <i_2 < i_3\leq 6} c_{i_1,i_2,i_3}\,|i_1,i_2,i_3\rangle
\end{equation}
only those Slater determinants $|i_1,i_2,i_3\rangle$ can show up which have exactly one index $i_k$ in each of the three sets $\{1,6\}$, $\{2,5\}$ and $\{3,4\}$. There are only $2^3=8$ such Slater determinants: $|1,2,3\rangle$, $|1,2,4\rangle$, $|1,3,5\rangle$, $|1,4,5\rangle$, $|2,3,6\rangle$, $|2,4,6\rangle$, $|3,5,6\rangle$ and $|4,5,6\rangle$. This universal statement is not in contradiction to the dimension $\binom{6}{3}=20$ of the $3$-fermion Hilbert space $\wedge^3[\mathcal{H}_1^{(6)}]$ since the $1$-particle states $|k\rangle$ depend on $|\Psi_3\rangle$.
If in addition $\vec{\lambda}$ is pinned to the facet described by saturation of (\ref{d=6b}) Eq.~(\ref{selec}) implies that
\begin{equation}\label{DBPinningstructure}
|\Psi_3\rangle = \alpha |1,2,3\rangle +\beta |1,4,5\rangle + \gamma |2,4,6\rangle\,.
\end{equation}
The three coefficients are free but should be chosen such that $\lambda_1 \geq \lambda_2 \geq \ldots\geq \lambda_6$.

We summarize these insights by
\begin{remark}\label{rem:Pinningisrelev}
Pinning corresponds to specific and simplified structures of the corresponding $N$-fermion quantum state $|\Psi_N\rangle$. In that sense pinning is highly physically relevant. It is also remarkable that pinning as phenomenon in the elementary $1$-particle picture allows to reconstruct the structure of $|\Psi_N\rangle$ as object in the important $N$-particle picture.
\end{remark}

In \cite{SuplMat} strong evidence is provided that the structural implications of exact pinning also hold (approximately) for quasipinning.
Due to this stability quasipinning is highly physical relevant.

A first physical application was suggested in \cite{CS2013} in form of a variational method. If
quasipinning for few-fermion ground states $|\Psi_N\rangle$ turns out to be generic for
not too strong interactions one can use the structural insights and choose an ansatz
for the ground state based on (\ref{selec}). By minimizing the energy expectation value w.r.t~the
corresponding non-zero coefficients $c_{\textbf{i}}$ and NO $|i\rangle$ one would obtain a good approximation
to the unknown exact ground state. Such an ansatz, a linear combination of several (up to 10'000) Slater determinants,
is well-known in quantum chemistry as multi-configurational self-consistent field (MCSCF) method.
However, for our variational optimization we would choose just a few, but very carefully chosen Slater determinants.

\section*{Acknowledgments}
We thank M.Christandl for helpful discussions and acknowledge financial support from the
Swiss State Secretariat for Education and Research supporting COST action MP1006 and 
from the Swiss National Science Foundation (Grant P2EZP2 152190).

\bibliographystyle{ws-procs975x65}
\bibliography{bibliography}

\end{document}